\documentclass{iau}

\usepackage{amsmath}
\usepackage{graphicx}
\usepackage{multirow}

\newcommand{\mnras}{MNRAS}
\newcommand{\apj}{ApJ}

\newcommand{\nat}{Nature}
\newcommand{\aap}{A\&A}
\newcommand{\pasp}{PASP}
\newcommand{\solphys}{SolPhys}

\begin{document}

\lefttitle{M. Leitzinger et al.}
\righttitle{Spectroscopic investigation of the southern late-K/early-M dwarf binary CC Eri}

\jnlPage{1}{7}
\jnlDoiYr{2021}
\doival{10.1017/xxxxx}

\aopheadtitle{Proceedings IAU Symposium}
\editors{C. Sterken,  J. Hearnshaw \&  D. Valls-Gabaud, eds.}

\title{Spectroscopic investigation of the southern late-K/early-M dwarf binary CC Eri}

\author{M. Leitzinger$^{1}$, P. Odert$^{1}$, R. Greimel$^{2}$, P. Kab\'ath$^{3}$, J. Lipt\'ak$^{3,4}$, R. Karjalainen$^{3}$,\\ P. Heinzel$^{3,5}$, J. Wollmann$^{3}$, E.W. Guenther$^{6}$}
\affiliation{$^{1}$Institute of Physics/AGP, University of Graz, Universitätsplatz 5, 8010 Graz, Austria\\
$^{2}$RG Science, Schanzelgasse 17, 8010 Graz, Austria\\
$^{3}$Astronomical Institute, The Czech Academy of Sciences, 25165 Ondřejov, Czech Republic\\
$^{4}$Astronomical Institute of Charles University, V Hole\'ovi\'ck\a'ch 2, 180 00 Prague, Czech Republic\\
$^{5}$University of Wrocław, Center of Scientific Excellence – Solar and Stellar Activity, 
Kopernika 11, 51-622 Wrocław, Poland\\
$^{6}$Thüringer Landessternwarte, TLS, Sternwarte 5, 07778 Tautenburg, Germany}

\begin{abstract}
In the recent past investigations of stellar Coronal Mass Ejections (CMEs) and flares/superflares gained special attention because of their relevance for planetary habitability
and stellar evolution. With respect to that, we present an observing campaign at the ESO152 telescope on LaSilla operated by the PLATOSpec consortium of the
late-K/early-M dwarf binary CC Eri. The ESO152 telescope is currently equipped with an Echelle spectrograph which enables the investigation of a broad spectral range,
allowing the investigation of CMEs and flares/superflares in various spectral lines. Within the observing campaign CC Eri was not observed by TESS, but we have
coordinated g'-band photometry to cross-check for white light flares. Preliminary results of the observing campaign are presented.
\end{abstract}


\maketitle

\section{Introduction}

Superflares, i.e. very energetic flares defined by an energy threshold of 10$^{33}$~erg \citep[see e.g.][]{Schaefer2000}, have been statistically identified even on slow rotating solar-like stars a decade ago \citep{Maehara2012}. Spectroscopic observations of superflares are rare. \citet{Hawley1991} present spectroscopic observations of a superflare on the dMe star AD~Leo and only recently superflares have been observed in H$\alpha$ on solar-like stars \citep{Namekata2021, Namekata2024, Leitzinger2024}. Here we present dedicated optical spectroscopic monitoring of the young and active southern binary CC~Eri with the main goal to spectroscopically characterize superflares. CC~Eri has been selected as a target 
\begin{figure}[h!]
    \includegraphics[scale=.22]{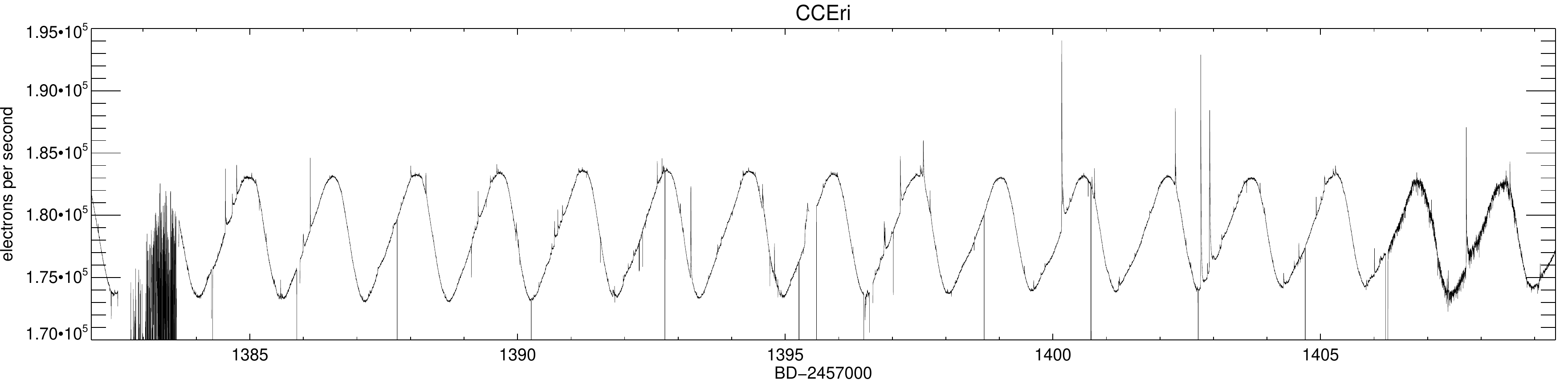}
        \caption{TESS light curve (sector 3) of CC~Eri. One can already identify by eye few stronger and numerous smaller flares.}
    \label{TESSlight}
\end{figure}
because it shows one of the largest flare/superflare rates in the southern hemisphere (see Fig.~\ref{TESSlight}), as obtained from TESS data. CC~Eri is a dK7e+dM3e spectroscopic binary, located at a distance of 11.5~pc, with a logLx of 29.2 ... 29.8~erg~s$^{-1}$, an orbital period of 1.56~d, an inclination of 40 ... 50$^{\circ}$, and a mass ratio of 2 \citep[see][]{Amado2000, Karmakar2017}. The expected flare rate of CC~Eri is 10 ... 39~d$^{-1}$ (E$_{flare}>$10$^{32}$~erg) for the XUV \citep[estimated based on][]{Audard2000}, 0.6 ... 2~d$^{-1}$ for H$\alpha$ \citep[estimated based on][]{Leitzinger2020}, and 4.3~d$^{-1}$ derived from TESS data.

\section{Observations}

CC~Eri was monitored from Sept. 2023 to Jan. 2024 with the ESO1.52m (ESO152) telescope on LaSilla, Chile hosted by the PLATOSpec consortium\footnote{\url{https://stel.asu.cas.cz/plato/}} consisting of the Astronomical Institute, Ondrejov of the Czech Academy of Sciences, Czech Republic, the Landessternwarte Tautenburg, Germany and the Universidad Catolica, Chile. PLATOSpec will perform spectroscopic follow-up observations (Echelle spectroscopy with R=70000) of PLATO targets. Before the installation of PLATOSpec at the ESO152, PUCHEROS+ \citep[updated version of PUCHEROS;][]{Vanzi2012}, an Echelle spectrograph with R$\sim$18000 is installed. In addition, the ESO152 provides two finder telescopes each with 15cm apertures. OndCam, a CMOS camera with Sloan filters and GrazCam, also a CMOS camera with Sloan and Johnson filters as well as a transmission grating for low-resolution spectroscopy, are installed on the finder telescopes, for coordinated observations. Spectroscopic data of CC~Eri were collected in more than 60 nights, or more than 2300 spectra, summing up to a total on-source time of close to 200 hours. For the majority of nights coordinated g'-band photometry is available, whereas only few nights are available with coordinated low-resolution spectroscopy. The spectra have been reduced and extracted with the CERES+ pipeline, an updated version of the CERES pipeline \citep{Brahm2017}. The photometric data have been reduced with a custom written pipeline \citep{Fryda2023} and light curves have been extracted using aperture photometry.

\section{Results}
\begin{figure}[h!]
    \includegraphics[scale=.58]{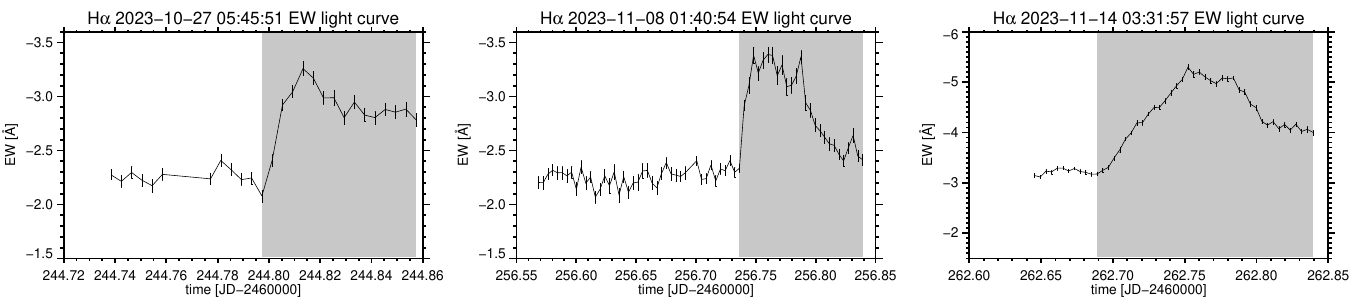}
        \caption{EW H$\alpha$ light curves of the three strongest flares found in the spectroscopic monitoring data of CC~Eri. The right most light curve shows the largest flare. The grey shaded area highlights the   
                 flare.}
    \label{EWlight}
\end{figure}

To identify flares we have two possibilities, either we use the coordinated photometry or to derive light curves from the spectra. As we do not have coordinated photometry for every night and not every flare may have a detectable white light component, we identify flares from light curves derived from chromospheric spectral lines, such as the Balmer lines together with temperature sensitive lines such as the HeI (5876\AA) line. We derive the equivalent width (EW) for every spectrum and line and plot then the EW time series to identify flares and superflares. In Fig.~\ref{EWlight} we show the three strongest flares which we found in the spectroscopic time series data of CC~Eri. In a first search we identified 18 flares. Only some flares have been fully captured, for some we have captured either the impulsive phase, or the gradual phase only.
The three selected flares reveal H$\alpha$ energies in the range of 2-8$\times$10$^{32}$~erg, which is for the least energetic flare roughly one order of magnitude below the superflare threshold of 10$^{33}$~erg. For one flare, namely for flare 2023-11-08 we estimate the g'-band energy which makes a much 
\begin{figure}
    \includegraphics[scale=.26]{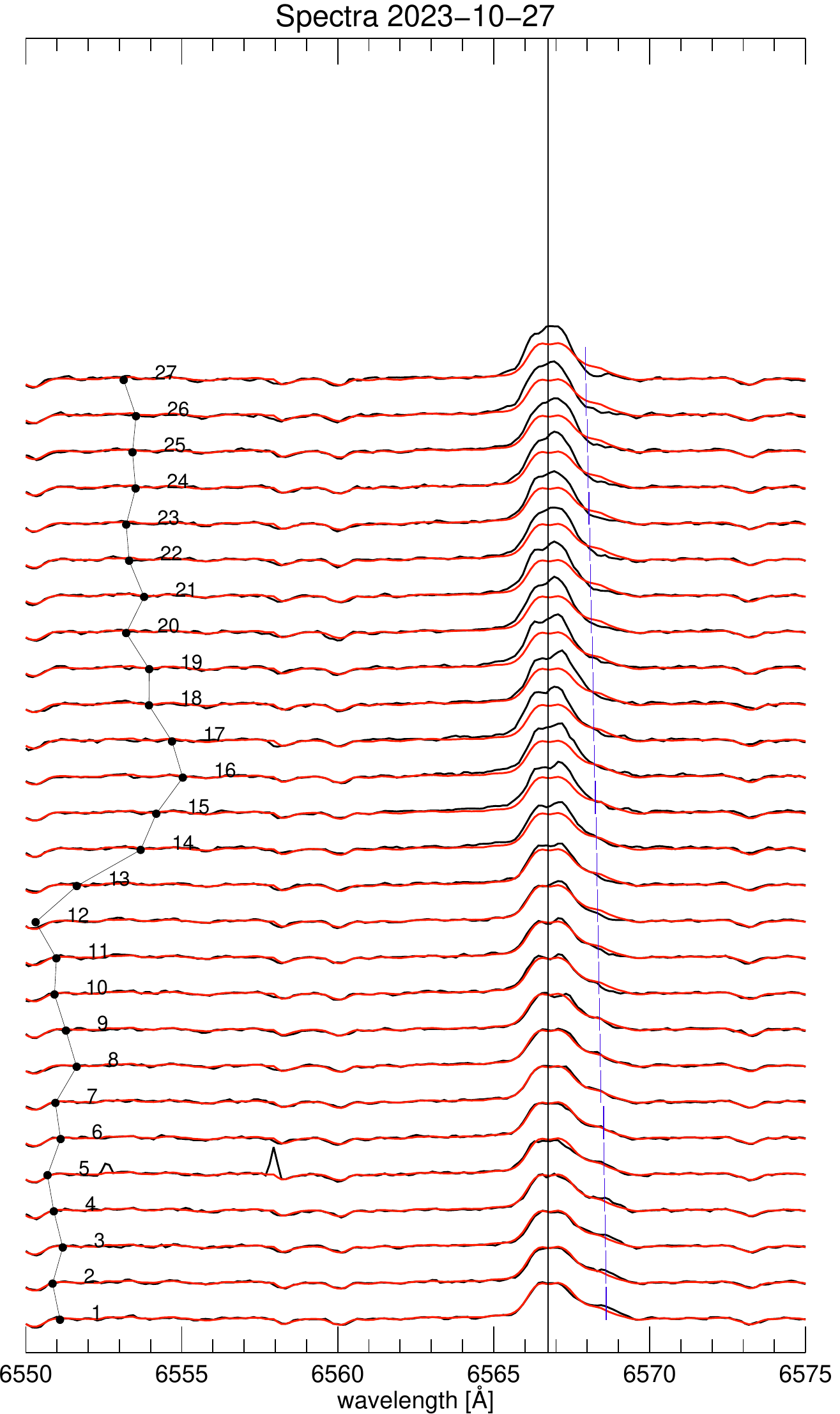}
    \includegraphics[scale=.26]{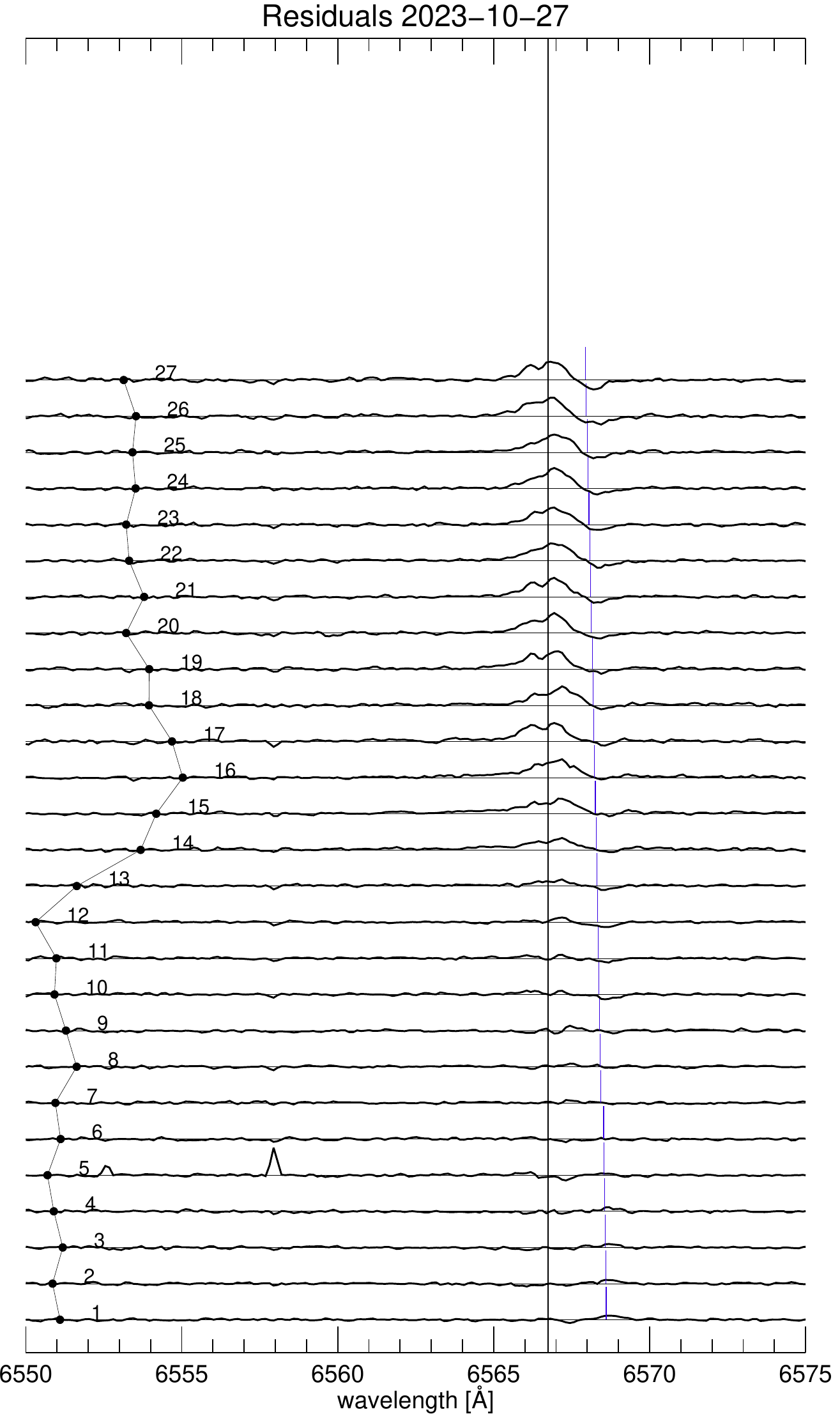}
    \includegraphics[scale=.26]{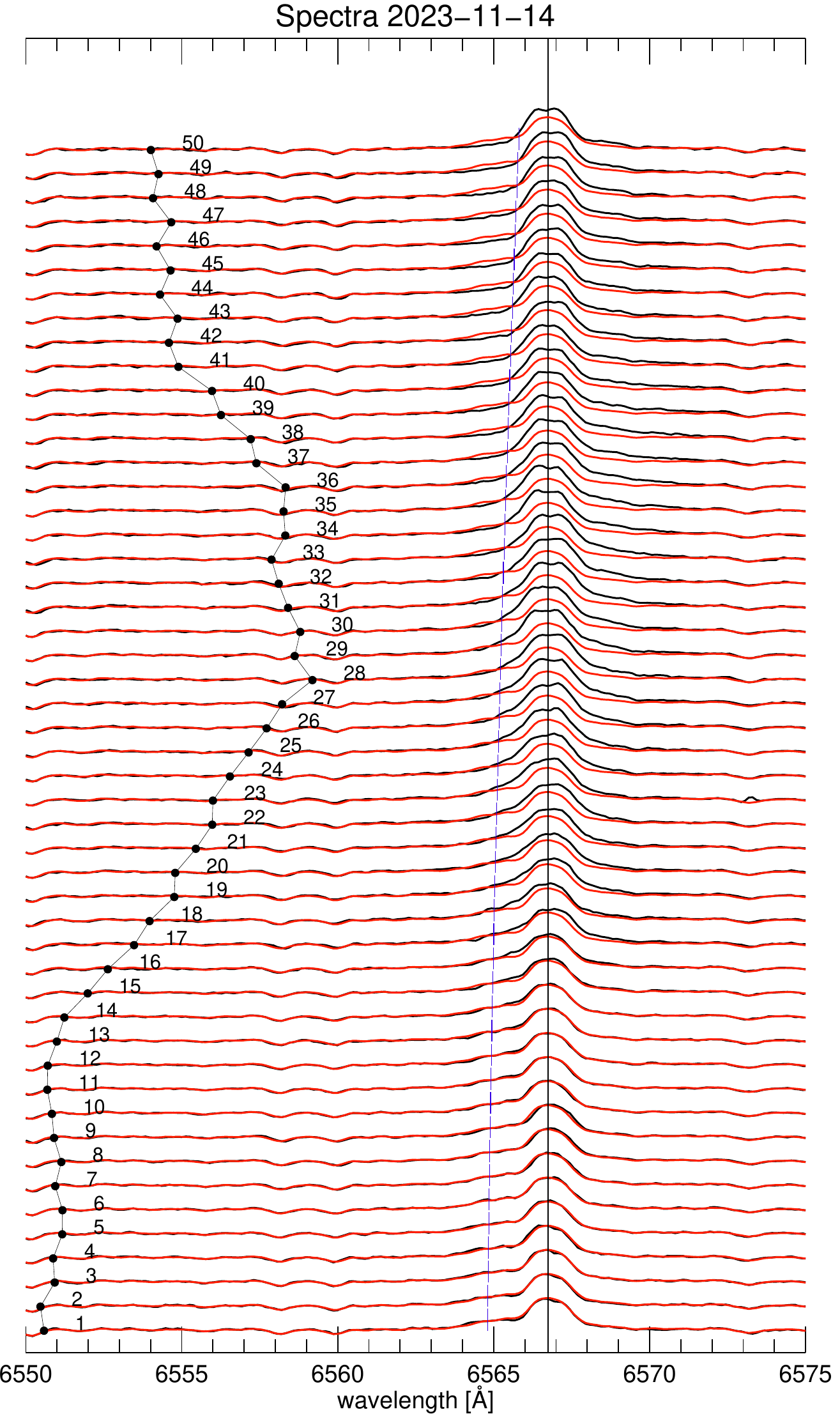}
    \hspace*{1.7cm}
    \includegraphics[scale=.26]{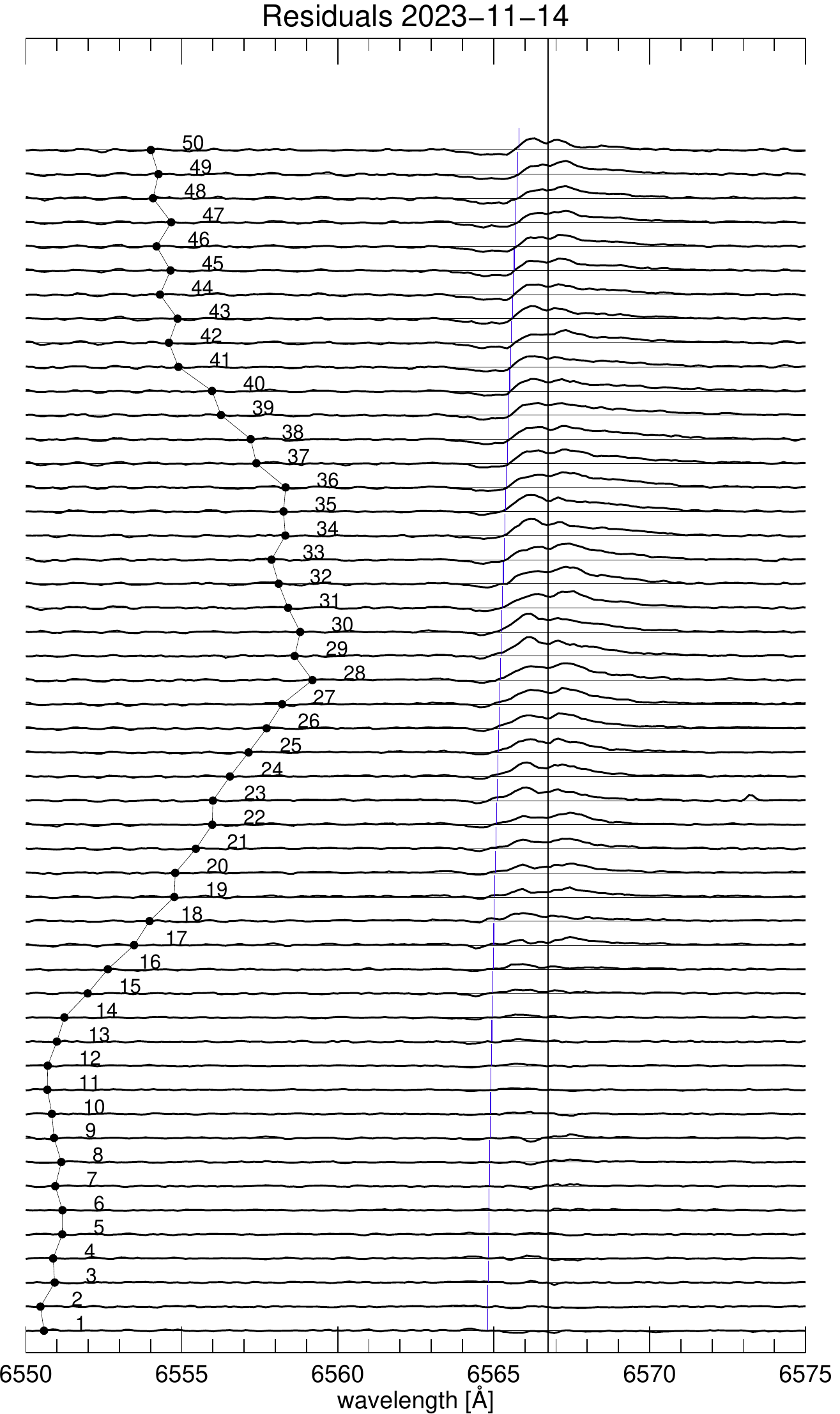}
        \caption{Left column: H$\alpha$ spectral time series (black solid line) of the flares 2023-10-27 and 2023-11-14 overplotted with the corresponding pre-flare spectrum (red solid line). Overplotted is the EW H$\alpha$ light curve (numbered data points) to identify impulsive and gradual flare phases. The blue vertical lines indicate the position of the H$\alpha$ profile of the dMe component of the CC~Eri system. The black solid vertical line denotes the H$\alpha$ line core of the primary dKe component. Right column: Corresponding residual spectral H$\alpha$ time series.   }
    \label{flareasym}
\end{figure}
larger portion of the white-light energy especially if compared to spectral line energies, such as e.g. H$\alpha$. The g'-band energy for the flare is above the 10$^{33}$~erg threshold. As CC~Eri is a binary one needs to consider for the computation of flare energies the quiet flux level of the flaring component. However, for the calculation of the g'-band flare energies we use the magnitude of the binary and not
of the components as those are not available. Considering this, the g'-band energy is an upper limit. In Table~\ref{flaretab} we list the flare energies (H$\alpha$, g'-band) and the flare durations.
\begin{table}
 \begin{center}
 \caption{Flare characteristics of the flares shown in Fig.~\ref{EWlight}.}\label{flaretab}
 {\tablefont\begin{tabular}{@{\extracolsep{\fill}}cccccc}
    \midrule
              &     E$_{g'-band}$       &      E$_{H\alpha}$       & total duration & duration impulsive phase & duration gradual phase\\
              &          [erg]          &            [erg]         &      [min]     &           [min]          &          [min]        \\
   2023-10-27 &           -             & 2$\times$ 10$^{32}$      &     $>$45      &           12             &          $>$33         \\   
   2023-11-08 &   4$\times$ 10$^{33}$   & 4$\times$ 10$^{32}$      &       78       &            9             &             69         \\
   2023-11-14 &           -             & 8$\times$ 10$^{32}$      &    $>$114      &           48             &         $>$ 66          \\ 
   
    \midrule
    \end{tabular}}
    \end{center}
\end{table}
Two of the energetic flares/superflares reveal line wing asymmetries in H$\alpha$, those are 2023-10-27 and 2023-11-14. In Fig.~\ref{flareasym} we show the spectral time series of both events (left column) together with their spectral residual H$\alpha$ time series (right column) for better visibility of the line asymmetries. Flare 2023-10-27 reveals a blue extra emission in the impulsive flare phase, while flare 2023-11-14 reveals a red extra emission in the gradual flare phase.

 \section{Discussion and Outlook}

We present preliminary results of an observing campaign aiming at the search for and the spectral characterization of superflares. CC~Eri is an excellent target for this purpose as the binary shows numerous flares and also a few superflares. However, it is surprising that we do not see significant evidence of signatures of erupting filaments/prominences, although the target flares frequently and we have captured so far 18 flares spectroscopically. The analysis of asymmetries is hampered by the fact that the components need to be separated to analyze each of their line profiles. Up to now we only studied the wings of the lines opposite to the location of the secondary component. For asymmetries not overlapping with the secondary component we are able to perform an analysis of spectral line asymmetries and extra emissions/absorptions. This is the case for the two presented superflares in Fig.\ref{flareasym}.

In the impulsive phase of flare 2023-10-27 starting at spectrum 14, a blue wing extra emission forms which lasts until spectrum 17 (one spectrum after the flare peak). Such blue wing asymmetries in the impulsive phase of flares are usually interpreted as chromospheric evaporation, if seen in hotter lines. During solar H$\alpha$ flares gentle chromospheric evaporation is known to occur which is very slow and reaches few km~s$^{-1}$ \citep[see e.g.][]{Schmieder1987}. \citet{Heinzel1994} discuss also blue asymmetries in the impulsive phase of solar flares and conclude that those are very short-lived (in the order of tens of seconds) and may stem from a small part of the flare. On the other hand blue extra emissions/absorptions in Balmer lines are also the signature of erupting prominences/filaments. But if the projected velocity of the signature is low, one can not unambiguously distinguish from flare related plasma motions \citep{Houdebine1990, Gunn1994}. In flare 2023-11-14 we see a long-lived red asymmetry in the gradual phase which arises in spectrum 32 and is visible until spectrum 41. In solar flares, red asymmetries are seen in the gradual flare phase as down-flowing plasma in post-flare loops also termed coronal rain. Only recently similar red asymmetries on the active dMe star AD~Leo were presented by \citet{Wollmann2023} who interpreted those as coronal rain.

For stellar spectroscopic observations one has to keep two important things in mind. First, stellar signals are always disk-integrated, and second, line asymmetries in stellar spectra are always seen in projection. So even if, as reported in \citet{Heinzel1994}, blue asymmetries stem from small parts of the flare, this may probably not or only hardly affect a disk integrated spectrum, even though on young and active stars, active regions and thus related flare areas may be larger than on the present-day Sun. Moreover, we usually do not know the orientation of the flare loops. Therefore, when interpreting stellar observations, this needs to be taken into account. 
 
One main aim of this study is the spectral characterization of supeflares which is work in progress and has been so far only done on a simple comparative basis of quiet and flare peak spectrum for the event 2023-11-14. We see the rise of Balmer lines H$\alpha$, H$\beta$, H$\gamma$, H$\delta$, the cores of the sodium lines NaID2+D1, the temperature sensitive line HeI (5876\AA{}), the MgI triplet, several FeI, FeII, CaI, and CrI spectral lines, and few lines which still need to be identified. We did not detect the also temperature sensitive line HeII (4686 \AA) which ususally appears in hot plasma \citep[see e.g.][]{Muheki2020b}. Moreover, a comparative spectral analysis of flares and superflares will be performed revealing possible differences between these two classes of flares.
\\
\\
\noindent\textbf{Acknowledgements}\\
The authors acknowledge the Austrian Science Fund (FWF): 10.55776/I5711 and the Czech Science Foundation (GACR): 22-30516K. PH was supported by the program ’Excellence Initiative - Research University’ for years 2020-2026 at University of Wrocław, project No. BPIDUB.4610.96.2021.KG.


\begin{thebibliography}{}

\bibitem[{Amado} et~al., 2000]{Amado2000}
{Amado}, P.~J., {Doyle}, J.~G., {Byrne}, P.~B.et al. 2000, \newblock {\em \aap}, 359, 159--167.

\bibitem[{Audard} et~al., 2000]{Audard2000}
{Audard}, M., {G{\"u}del}, M., {Drake}, J.~J. et al. 2000, \newblock {\em \apj}, 541, 396--409.

\bibitem[{Brahm} et~al., 2017]{Brahm2017}
{Brahm}, R., {Jord{\'a}n}, A., \& {Espinoza}, N. 2017, \newblock {\em \pasp}, 129(973), 034002.

\bibitem[{Fryda}, 2023]{Fryda2023}
{Fryda}, J.. 2023, Master Thesis, Charles University, Prague, Czech Republic

\bibitem[{Gunn} et~al., 1994]{Gunn1994}
{Gunn}, A.~G., {Doyle}, J.~G., {Mathioudakis}, M. et al. 1994, \newblock {\em \aap}, 285, 489--496.

\bibitem[{Hawley} and {Pettersen}, 1991]{Hawley1991}
{Hawley}, S.~L. \& {Pettersen}, B.~R. 1991, \newblock {\em \apj}, 378, 725.

\bibitem[{Heinzel} et~al., 1994]{Heinzel1994}
{Heinzel}, P., {Karlicky}, M., {Kotrc}, P. et al., 1994, \newblock {\em \solphys}, 152, 393--408.

\bibitem[{Houdebine} et~al., 1990]{Houdebine1990}
{Houdebine}, E.~R., {Foing}, B.~H., \& {Rodono}, M. 1990, \newblock {\em \aap}, 238, 249--255.

\bibitem[{Karmakar} et~al., 2017]{Karmakar2017}
{Karmakar}, S., {Pandey}, J.~C., {Airapetian}, V.~S. et al. 2017, \newblock {\em \apj}, 840(2), 102.

\bibitem[{Leitzinger} et~al., 2020]{Leitzinger2020}
{Leitzinger}, M., {Odert}, P., {Greimel}, R. et al. 2020, \newblock {\em \mnras}, 493(3), 4570--4589.

\bibitem[{Leitzinger} et~al., 2024]{Leitzinger2024}
{Leitzinger}, M., {Odert}, P., \& {Greimel}, R. 2024, \newblock {\em \mnras}, 532(2), 1486-1503.

\bibitem[{Maehara} et~al., 2012]{Maehara2012}
{Maehara}, H., {Shibayama}, T., {Notsu}, S. et al. 2012, \newblock {\em \nat}, 485, 478--481.

\bibitem[{Muheki} et~al., 2020]{Muheki2020b}
{Muheki}, P., {Guenther}, E.~W., {Mutabazi}, T. et al. 2020, \newblock {\em \mnras}, 499(4), 5047--5058.

\bibitem[{Namekata} et~al., 2024]{Namekata2024}
{Namekata}, K., {Airapetian}, V.~S., {Petit}, P. et al. 2024, \newblock {\em \apj}, 961(1), 23.

\bibitem[{Namekata} et~al., 2021]{Namekata2021}
{Namekata}, K., {Maehara}, H., {Honda}, S. et al. 2021, \newblock {\em Nature Astronomy}, 6, 241--248.

\bibitem[{Schaefer} et~al., 2000]{Schaefer2000}
{Schaefer}, B.~E., {King}, J.~R., \& {Deliyannis}, C.~P. 2000, \newblock {\em \apj}, 529(2), 1026--1030.

\bibitem[{Schmieder} et~al., 1987]{Schmieder1987}
{Schmieder}, B., {Forbes}, T.~G., {Malherbe}, J.~M. et al. 1987, \newblock {\em \apj}, 317, 956--963.

\bibitem[{Vanzi} et~al., 2012]{Vanzi2012}
{Vanzi}, L., {Chacon}, J., {Helminiak}, K.~G. et al. 2012, \newblock {\em \mnras}, 424(4), 2770--2777.

\bibitem[{Wollmann} et~al., 2023]{Wollmann2023}
{Wollmann}, J., {Heinzel}, P., \& {Kab{\'a}th}, P. 2023, \newblock {\em \aap}, 669, A118.

\end{thebibliography}
\end{document}